\documentclass[journal]{IEEEtran}

\ifCLASSINFOpdf

\else

\fi

\usepackage{graphicx}
\usepackage{color}
\usepackage{subfigure}

\begin{document}

\title{Stochastic Geometry Analysis of Multi-Antenna Two-Tier Cellular Networks}

\author{Zheng Chen,~Ling~Qiu,~\IEEEmembership{Member,~IEEE},~and~Xiaowen~Liang
\thanks{The research has been supported by the 863 Program 2014AA01A702. The authors are with Key Laboratory of Wireless-Optical Communications, Chinese Academy of Sciences, School of Information Science and Technology, University of Science and Technology of China. Address: No. 96 Jinzhai Road, Hefei, Anhui Province, 230026, P. R. China. L. Qiu is the corresponding author (e-mail: lqiu@ustc.edu.cn).}}

\markboth{}%
{Shell \MakeLowercase{\textit{et al.}}: Bare Demo of IEEEtran.cls for Journals}

\maketitle

\begin{abstract}
In this paper, we study the key properties of multi-antenna two-tier networks under different system configurations. Based on stochastic geometry, we derive the expressions and approximations for the users' average data rate. Through the more tractable approximations, the theoretical analysis can be greatly simplified. We find that the differences in density and transmit power between two tiers, together with range expansion bias significantly affect the users' data rate.  Besides, for the purpose of area spectral efficiency (ASE) maximization, we find that the optimal number of active users for each tier is approximately fixed portion of the sum of the number of antennas plus one. Interestingly, the optimal settings are insensitive to different configurations between two tiers. Last but not the least, if the number of antennas of macro base stations (MBSs) is sufficiently larger than that of small cell base stations (SBSs), we find that range expansion will improve ASE.
\end{abstract}

\begin{IEEEkeywords}
Area spectral efficiency, heterogeneous networks,  multi-antenna base stations, stochastic geometry.
\end{IEEEkeywords}

\IEEEpeerreviewmaketitle

\section{Introduction}

\IEEEPARstart{H}{eterogeneous} networks (HetNets), which is composed of MBSs overlaid with SBSs,  has been recognized as a promising approach to cope with the 1000x traffic demand \cite{dense:bushan}.  However, the differences in transmit power and BS density between MBSs and  SBSs make  the properties of HetNets different from the traditional single-tier networks.

To characterize the performance of HetNets, stochastic geometry has been identified as a powerful tool  because of its accuracy and tractability. By modeling MBSs and SBSs as independent Poisson point processes (PPPs), the performance of HetNets has been extensively studied in \cite{hetnet:Mukherjee}\cite{hetnet:dhillon}\cite{hetnet:jo}. Especially, the technique of cell range expansion for load-balancing was considered in \cite{hetnet:jo}. It is found that, although range expansion improves the fairness between users, the overall throughput will be degraded. However, \cite{hetnet:Mukherjee}\cite{hetnet:dhillon}\cite{hetnet:jo} only considered single-antenna BSs, the key properties of multi-antenna HetNets cannot be revealed. In  multi-antenna single-tier networks, the ASE and energy efficiency have been studied in \cite{li:eemimo}\cite{chen:twc}. \cite{Eid:renzo} analyzed the error probability of single-tier Poisson distributed networks. However, the interplay between different tiers in HetNets cannot be understood in \cite{li:eemimo}\cite{chen:twc}\cite{Eid:renzo}. To take a step further, the coverage probability of multi-antenna HetNets was investigated in \cite{dhillon:mimohetnet}\cite{li:sdma}\cite{mimohetnet:gupta}.  \cite{diversity:tanbourgi} studied the coverage probability of multi-antenna HetNets with transmit-receive diversity. However, \cite{dhillon:mimohetnet}\cite{li:sdma}\cite{mimohetnet:gupta}\cite{diversity:tanbourgi} focused more on the analysis of signal-to-interference-plus-noise-ratio (SINR), which is usually in complex form. It is difficult to analyze the properties of multi-antenna HetNets under different system parameters theoretically based on the previous works. As a results, the impact of differences between two tiers  on system performance have not been well investigated.

In this paper, we derive the expressions and tight approximations for users' average data rate in multi-antenna two-tier networks. Through the more tractable approximations, we study how different system parameter configurations affect the users' average data rate theoretically. Furthermore, also based on the approximations, we obtain the optimal number of active users of each tier to maximize ASE. We study the optimal settings under different system parameters. We find that the ratio of BS density and transmit power $\frac{\lambda_s}{\lambda_m},\frac{P_s}{P_m}$ and the range expansion bias mainly affect the users' data rate and the optimal settings.\footnote{$\frac{\lambda_s}{\lambda_m},\frac{P_s}{P_m}$ are the ratios of BS density, transmit power between MBSs and SBSs, which will be defined in the main body of this paper.}  Moreover, the results based on the approximations are validated through numerical results. The novel and insightful findings of this paper as follows:
\begin{itemize}
  \item The average data rate of macro or small cell users is insensitive to the number of active users of the other tier. If range expansion is considered, the users' data rates of both tier increase with $\frac{\lambda_s}{\lambda_m},\frac{P_s}{P_m}$. However, if range expansion is not considered, the users' data rate is insensitive to $\frac{\lambda_s}{\lambda_m},\frac{P_s}{P_m}$.
  \item For each tier, the ratio between the optimal number of active users and the sum of the number of antennas plus one nearly remains fixed, which is insensitive to different system configurations. With the number of active users set as the optimal number, ASE increases linearly with the number of antennas of both tier.
  \item If the number of antennas of MBSs is sufficiently larger than that of SBSs, range expansion will improve ASE. Besides,  the region of the number of antennas  where range expansion improves ASE is insensitive to $\frac{\lambda_s}{\lambda_m}$. However, larger $\frac{P_s}{P_m}$ will expand the improvement region.
\end{itemize}

\section{System Model}
We consider a downlink two-tier network where the macro tier is overlaid with co-channel deployed SBSs. The positions of MBSs and SBSs are modeled as  PPPs \(\Phi_m,\Phi_s\) with density $\lambda_m,\lambda_s$.  The MBSs and SBSs have transmit power $P_m,P_s$ and are equipped with $M_m,M_s$ antennas. The single-antenna users are located according to some stationary point process, which is independent of \(\Phi_m,\Phi_s\).  We adopt standard path loss propagation model with path loss exponent  $\alpha>2$. User association is based on the long-term average biased received power. The range expansion bias  for SBSs is $B$.

Each MBS and SBS serves $K_m,K_s$ active users at each slot. We  consider zero-forcing (ZF) precoding and assume perfect channel state information at each BS. Thus, we have $K_m \le M_m,K_s \le M_s$. We follow the infinite user density assumption in \cite{dhillon:mimohetnet}\cite{li:sdma}. Therefore, there are at least $K_m,K_s$ users covered by each MBS and SBS.  The transmit power is allocated equally for the active users in each cell. The small scale fading on each link is i.i.d. Rayleigh fading. The thermal noise is ignored in this paper.  Without loss of generality, we will focus on the analysis of a typical user located at the origin \cite{book:baccelli}. The signal-to-noise-ratio (SIR) of the typical user located at the origin is
\begin{equation}
\begin{array}{l}
SIR_l = \frac{\frac{P_lg_{l0}}{K_l\|x_0\|^{\alpha}}}{\sum _{x_i\in \Phi_m \backslash \{x_0\}}\frac{P_mh_{mi}}{K_m\|x_i\|^{\alpha}}+\sum _{x_i\in \Phi_s \backslash \{x_0\}}\frac{P_sh_{si}}{K_s\|x_i\|^{\alpha}}},
\end{array}
\end{equation}
where $x_0$ indicates the BS the typical user associated with, and $l=m,s$ depends on which type BS $x_0$ is. According to \cite{dhillon:mimohetnet}\cite{li:sdma}, we know the equivalent channel gains $g_{l0} \sim Gamma(M_l+1-K_l,1)$, $h_{mi}\sim Gamma(K_m,1)$ and $h_{si}\sim Gamma(K_s,1)$.\footnote{Actually, the proposition $h_{mi},h_{si}$ are Gamma distributed is an accurate approximation, which is widely adopted in previous works \cite{chen:twc}\cite{dhillon:mimohetnet}\cite{mimohetnet:gupta}\cite{li:sdma}.  }
%\footnote{In fact, it is an approximation that $h_{mi},h_{si}$ are Gamma distributed. This approximation has been demonstrated to be quite accurate for the analysis of multi-antenna networks in \cite{chen:twc}.}

\section{Analysis of the Typical user}
In this section, we will study the average data rate of the typical user, especially, the impact of different configurations between two tiers will be discussed.

\subsection{Theoretical Analysis}
Taking average over spatial distribution and channel power distribution, we can obtain the average data rate of the typical user, which is provided in the following theorem.
\newtheorem{theorem}{\bf{Theorem}}
\begin{theorem}\label{thdatarate}The average data rate $R_m,R_s$ of the typical user in macro tier and small cell tier are given by
\begin{equation}\label{eqdatarate}
\begin{array}{l}
R_m=\int_0^\infty  {\frac{{\left( { 1 + \frac{\lambda _s}{\lambda _m} \left( {\frac{{P_s B}}{{P_m }}} \right)^{2/\alpha } } \right)\left( {1 - \left( {\frac{1}{{1 + z}}} \right)^{M_m  + 1 - K_m } } \right)}}{z\left({ F\left( {z,K_m } \right) + \frac{\lambda _s}{\lambda _m} \left( {\frac{{P_s B}}{{P_m }}} \right)^{2/\alpha } F\left( {\frac{{zK_m }}{{K_s B}},K_s } \right)}\right)}dz},\\
R_s= \int_0^\infty  {\frac{{\left( {\frac{\lambda _m}{\lambda _s} \left( {\frac{{P_m }}{{P_s B}}} \right)^{2/\alpha }  +1  } \right)\left( {1 - \left( {\frac{1}{{1 + z}}} \right)^{M_s  + 1 - K_s } } \right)}}{z\left({\frac{\lambda _m}{\lambda _s} \left( {\frac{{P_m }}{{P_s B}}} \right)^{2/\alpha } F\left( {\frac{{zK_s B}}{{K_m }},K_m } \right) + F\left( {z,K_s } \right)}\right)}dz},
\end{array}
\end{equation}
where $F\left( {x,y} \right) = 1 + x^{2/\alpha } \int_{x^{ - 2/\alpha } }^\infty  {\left( {1 - {{\left( {1 + u^{ - \alpha /2} } \right)^{-y} }}} \right)du}$.
\end{theorem}

\begin{IEEEproof}
The proof is given in the appendix.
\end{IEEEproof}

Theorem \ref{thdatarate} provide tractable expressions for the average data rate of users in multi-antenna two-tier  networks. However, $R_m,R_s$ have complicated relationships with other system parameters. To further study how the differences between two tiers impact on the network performance, we provide  more tractable approximations for $R_m,R_s$, which are demonstrated to be quite tight in our numerical results.

\begin{theorem}\label{thlowerbound}The approximations for  ${R}_m,{R}_s$  are
\begin{equation}\label{eqlowerbound}
\begin{array}{l}
\tilde{R}_m = \int_0^\infty  {\frac{1}{z}\frac{{\left( {1  + \frac{\lambda _s}{\lambda _m} (\frac{{P_s B}}{{P_m }})^{2/\alpha } } \right)\left( {1 - e^{ - z {\frac{{M_m +1  - K_m }}{{K_m }}} } } \right)}}{{ H\left( z \right) +  \frac{\lambda _s}{\lambda _m} \left( {\frac{{P_s B}}{{P_m }}} \right)^{2/\alpha } H\left( {\frac{z}{B}} \right)}}} dz,\\
\tilde{R}_s = \int_0^\infty  {\frac{1}{z}\frac{{\left( {\frac{\lambda _m}{\lambda_s} \left( {\frac{{P_m }}{{P_s B}}} \right)^{2/\alpha }  + 1 } \right)\left( {1 - e^{ - z{\frac{{M_s + 1 - K_s }}{{K_s }}} } } \right)}}{{\frac{\lambda _m}{\lambda_s}\left( {\frac{{P_m }}{{P_s B}}} \right)^{2/\alpha } H\left( {zB} \right) + H\left( z \right)}}} dz,
\end{array}
\end{equation}
where $H\left( x \right) = 1 + x^{2/\alpha } \int_{x^{ - 2/\alpha } }^\infty  {\left( {1 - \exp \left( { - u^{ - \alpha/2 } } \right)} \right)} du$.
\end{theorem}

\begin{IEEEproof}
The proof is given in the appendix.
\end{IEEEproof}

From Theorem \ref{thdatarate}, $R_m,R_s$ are unrelated to the number of antennas of the other tier. Interestingly, besides the number of antennas, $\tilde{R}_m,\tilde{R}_s$ do not depend on the number of active users of the other tier either.  In fact, numerical results show that the original expressions $R_m,R_s$ are also insensitive to  the number of  active users of the other tier, i.e.,  $R_m,R_s$ behave similarly to $\tilde{R}_m,\tilde{R}_s$. Moreover, based on Theorem \ref{thdatarate}, we find that the ratios $\frac{\lambda_s}{\lambda_m},\frac{P_s}{P_m}$ and $B$ mainly affect $\tilde{R}_m,\tilde{R}_s$.

\newtheorem{lemma}{\bf{Lemma}}
\begin{lemma}\label{lemmadatarate}When $B>1$, we have 1) $\tilde{R}_m$ increases with $\frac{\lambda_s}{\lambda_m}$, $\frac{P_s}{P_m}$, and $B$; 2) $\tilde{R}_s$ increases with $\frac{\lambda_s}{\lambda_m}$ and $\frac{P_s}{P_m}$, but decreases with $B$.
However, when $B=1$,  $\tilde{R}_m,\tilde{R}_s$ are unrelated to $\lambda_s,{\lambda_m}$, ${P_s},{P_m}$ and can be simplified as
$\begin{array}{l}
\tilde{R}_m = \int_0^\infty  {\frac{{ {1 - e^{ - z {\frac{{M_m  +1- K_m }}{{K_m }}} } } }}{{ zH\left( z \right) }}} dz,\tilde{R}_s = \int_0^\infty  {\frac{{ {1 - e^{ - z {\frac{{M_s  +1- K_s }}{{K_s }}} } } }}{{ zH\left( z \right) }}} dz
\end{array}$.
\end{lemma}

\begin{IEEEproof}
The proof is given in the appendix.
\end{IEEEproof}

It is a well-known fact that the average data rate increases with $B$ for macro users, but decreases with $B$ for small cell users. However, instead of numerical results, this property is demonstrated theoretically through the tight approximations in this paper.  Furthermore, it is interesting to point out that both $\tilde{R}_m,\tilde{R}_s$ increase with $\frac{\lambda_s}{\lambda_m}$ and $\frac{P_s}{P_m}$ when $B>1$, which has never been reported in previous works. That is to say, the more small cells or the higher transmit power of small cells, the average data rate of both tiers will be larger. Another interesting result is that the approximations of users' data rate do not depend on  BS density nor BS transmit power when $B=1$, which is similar to the single-antenna scenario \cite{hetnet:jo}. The properties in Lemma \ref{lemmadatarate} will be validated for the original expressions $R_m,R_s$ through numerical results.

\begin{figure}[!t]
\centering
\includegraphics[width=3.1in]{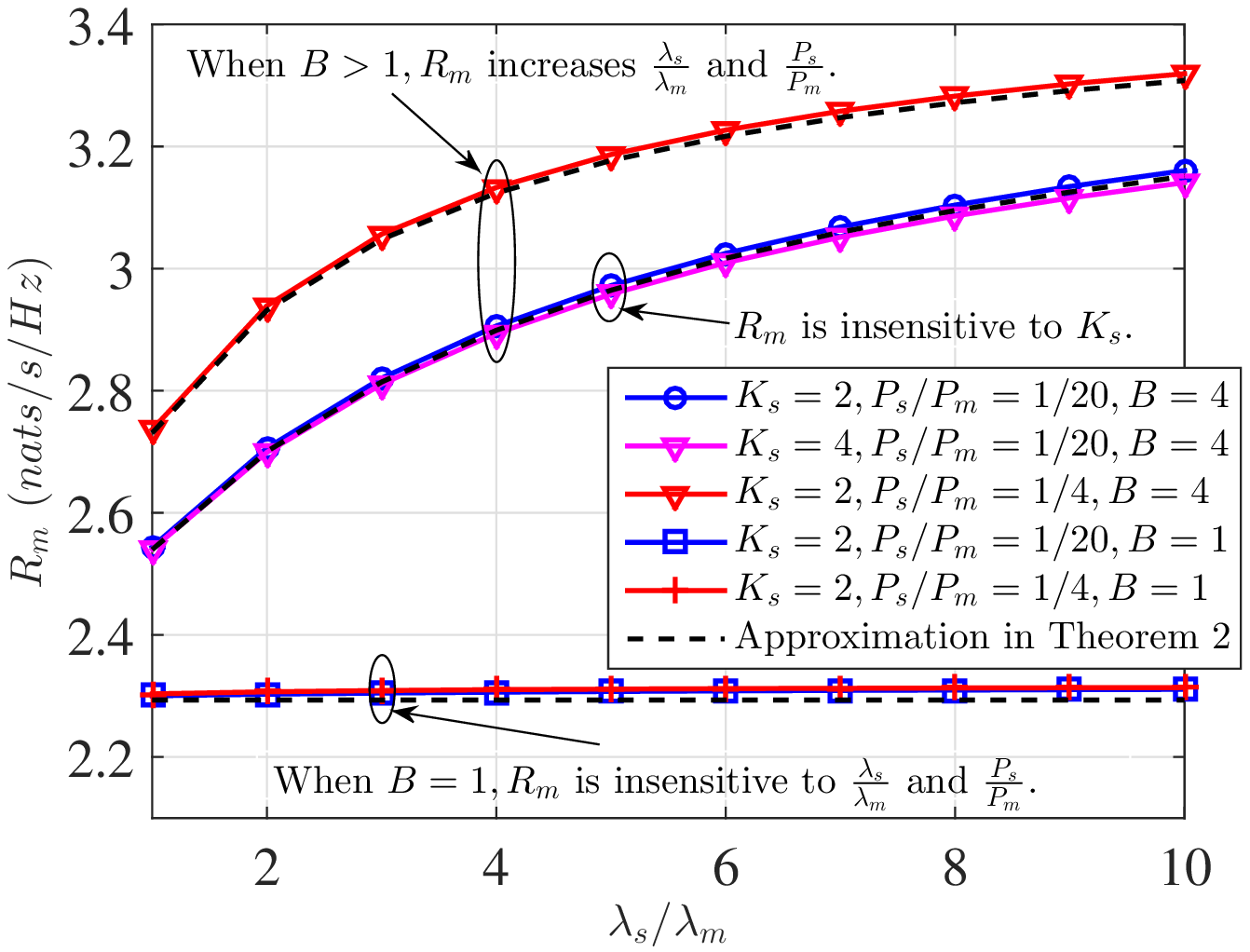} \\
\includegraphics[width=3.1in]{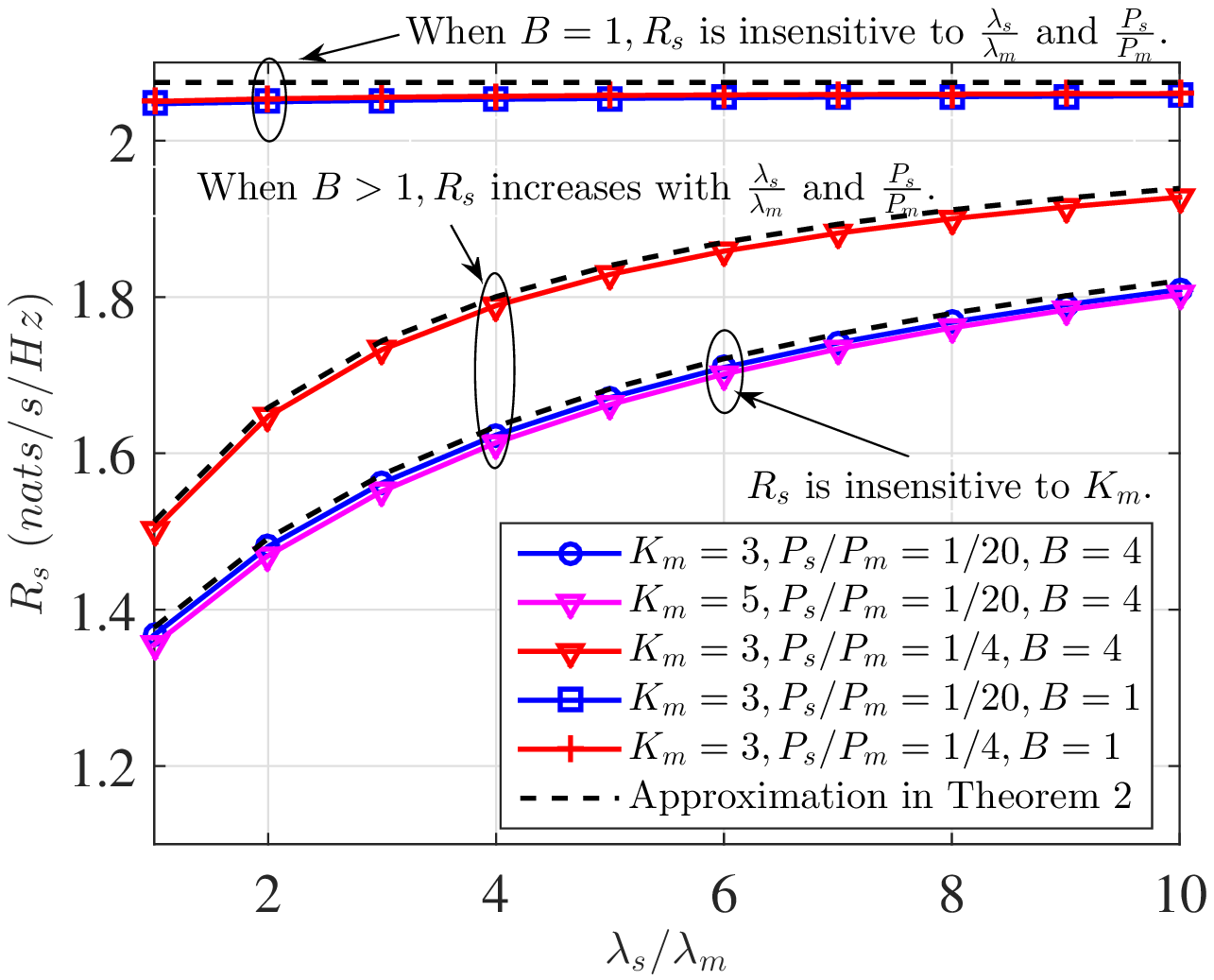}
\caption{$R_m,R_s$ with different system parameters. $\alpha=4$. In the figure $R_m$, we set $M_m=10,K_m=3$. In the figure $R_s$, we set $M_s=5,K_s=2$.}
\label{datarate}
\end{figure}

Fig. \ref{datarate} depicts $R_m,R_s$ with different system parameters. We can find that the approximations in Theorem \ref{thlowerbound} is quite accurate. Moreover, all the mentioned properties of $\tilde{R}_m,\tilde{R}_s$ are validated for $R_m,R_s$. Firstly, we find $R_m,R_s$ are insensitive  to the number of active users of the other tier. The reason is that, instead of $K_m,K_s$, it is the total powers $P_m,P_s$ which dominate  the cross-tier interference. Secondly, when $B>1$, we observe that both $R_m$ and $R_s$ increase with $\frac{\lambda_s}{\lambda_m}$, $\frac{P_s}{P_m}$. By increasing $\lambda_s$, we know that the inter-cell interference from the small cell tier will be stronger. However, the users will be closer to the associated BSs, which makes the desired signal power stronger. From Fig. \ref{datarate}, we know that the increase of desired signal power is larger than the increase of the interference power, which leads to the fact that $R_m,R_s$ increases with $\frac{\lambda_s}{\lambda_m}$. Similarly, for increasing $P_s$, we also know that the increase of desired power is larger than the interference power. Different from the case $B>1$, when $B=1$, $R_m,R_s$ are insensitive  to $\frac{\lambda_s}{\lambda_m}$ and $\frac{P_s}{P_m}$. That is to say, when $B=1$, the increase of desired signal power due to large $\lambda_s,P_s$ is counter-balanced by the increase of inter-cell interference. Comparing the results for $B=4$ and $B=1$, we find range expansion will improve $R_m$ but degrade $R_s$. This property is similar to the single-antenna HetNets \cite{hetnet:jo}. However, we consider a more general case where multi-antenna BSs are considered. Above all, all theoretical results based on the approximations $\tilde{R}_m,\tilde{R}_s$ are validated for $R_m,R_s$. The impact of differences between two tiers on users' data rate has been revealed.

\section{Area Spectral Efficiency Optimization}
In this section, we will study the optimal $K_m^*,K_s^*$ to maximize ASE. The key properties of $K_m^*,K_s^*$ under different system parameters will be revealed. How range expansion affects ASE will be discussed.

\subsection{The Optimal Number of Active Users}
Based on Theorem \ref{thdatarate}, the expression of ASE can be expressed as  $T=\lambda_mK_mR_m+\lambda_sK_sR_s$. We find that $K_m,K_s$  are  crucial system design parameters, which affect ASE of both tiers. Thus, we attempt to obtain the optimal $K_m^*,K_s^*$ to maximize ASE. It is difficult to maximize $T$ directly because of the complex expressions of $R_m,R_s$. Therefore, we obtain $\tilde{K}_m^*,\tilde{K}_s^*$ that maximize the approximations $\tilde{T}=\lambda_mK_m\tilde{R}_m+\lambda_sK_s\tilde{R}_s$ instead. Fortunately, due to the tightness of the approximations, the numerical results demonstrate that the results based on the $\tilde{T}$ are consistent with $T$.

In detail, we formulate the following problem,
\begin{equation}\label{pk}
\begin{array}{ll}
\max \limits_{K_m,K_s} &  \lambda_m K_m \tilde{R}_m + \lambda_sK_s \tilde{R}_s  \\
 s.t. & K_m\in\{0,1,...,M_m\}, K_s\in\{0,1,...,M_s\}.
\end{array}
\end{equation}

Substituting $u_m=\frac{K_m}{M_m+1},u_m=\frac{K_s}{M_s+1}$ into the above problem and relaxing $u_m,u_s$ to $[0,1]$, we arrive at the following problem,
\begin{equation}\label{pu}
\begin{array}{ll}
\max \limits_{u_m,u_s} &  \lambda_m  \left(M_m+1\right) G_m(u_m) + \lambda_s \left(M_s+1\right) G_s(u_s)  \\
 s.t. & u_m,u_s \in [0,1].   \\
\end{array}
\end{equation}
where ${G_m(u)} ={  \int_0^\infty  {\frac{u{\left( {1  + \frac{\lambda _s}{\lambda _m} (\frac{{P_s B}}{{P_m }})^{2/\alpha} } \right) \left( {1 - e^{ - z/u+z } } \right)}}{z\left({ H\left( z \right) +  \frac{\lambda _s}{\lambda _m} \left( {\frac{{P_s B}}{{P_m }}} \right)^{2/\alpha} H\left( {\frac{z}{B}} \right)}\right)}} dz}, G_s(u)={  \int_0^\infty  {\frac{{u\left( {\frac{\lambda _m}{\lambda_s} \left( {\frac{{P_m }}{{P_s B}}} \right)^{2/\alpha}  + 1 } \right) \left( {1 - e^{ - z/u +z} } \right)}}{z\left({\frac{\lambda _m}{\lambda_s}\left( {\frac{{P_m }}{{P_s B}}} \right)^{2/\alpha} H\left( {zB} \right) + H\left( z \right)}\right)}} dz}$. For the optimization problem $(\ref{pu})$, we have the following observations.

\begin{itemize}
  \item Firstly, $u_m,u_s$ can be optimized separately to maximize $G_m(u),G_s(u)$.
  \item Secondly, the optimal $u_m^*,u_s^*$ do not depend $M_m,M_s$, i.e., for arbitrary $M_m,M_s$, the optimal numbers of active users are $\left(M_m+1\right)u_m^*,\left(M_s+1\right)u_s^*$. \footnote{In fact, the number of active users should be rounded.}
  \item Furthermore, with the optimal $u_m^*,u_s^*$, the approximations of ASE $\tilde{T}^*=\lambda_m \left(M_m+1\right)G_m(u_m^*)+\lambda_s \left(M_s+1\right)G_s(u_s^*)$, which will increase linearly with $M_m,M_s$.
\end{itemize}

Through the second order derivatives, it is not difficult to show that both $G_m(u)$ and $G_s(u)$ are concave function of $u$. Therefore, the optimal $u_m^*,u_s^*$ can be obtained by setting first order derivative as zero.
\begin{theorem}\label{thu}$u_m^*$ is the solution of ${  \int_0^\infty  {\frac{1-e^{-z/u+z}-\frac{z}{u}e^{-z/u+z}}{z\left({ H\left( z \right) +  \frac{\lambda _s}{\lambda _m} \left( {\frac{{P_s B}}{{P_m }}} \right)^{{2}/{\alpha }} H\left( {\frac{z}{B}} \right)}\right)}} dz}=0$. $u_s^*$ is the solution of ${  \int_0^\infty  {\frac{1-e^{-z/u+z}-\frac{z}{u}e^{-z/u+z}}{z\left({\frac{\lambda _m}{\lambda_s}\left( {\frac{{P_m }}{{P_s B}}} \right)^{{2}/{\alpha }} H\left( {zB} \right) + H\left( z \right)}\right)}} dz}$. Both of the two equations have a unique solution located in $[0,1]$, which can be obtained through the bisection method.
\end{theorem}

\begin{IEEEproof}
We take $u_m$ as an example. Take the first order derivative $\frac{\partial}{\partial u}G_m(u)$, we arrive at the mentioned equation. $\frac{\partial}{\partial u}G_m(u)$ is a decreasing function of $u$. Besides, for the boundaries $0,1$, it is not difficult to show that  $\lim_{u\rightarrow0}\frac{\partial}{\partial u}G_m(u)={  \int_0^\infty  {\frac{1}{z\left({ H\left( z \right) +  \frac{\lambda _s}{\lambda _m} \left( {\frac{{P_s B}}{{P_m }}} \right)^{{2}/{\alpha }} H\left( {\frac{z}{B}} \right)}\right)}} dz}>0$ and   $\frac{\partial}{\partial u}G_m(1)={  -\int_0^\infty  {\frac{1}{\left({ H\left( z \right) +  \frac{\lambda _s}{\lambda _m} \left( {\frac{{P_s B}}{{P_m }}} \right)^{{2}/{\alpha }} H\left( {\frac{z}{B}} \right)}\right)}} dz}<0$. Therefore, there is a unique solution located in $[0,1]$, which can be solved through bisection method.
\end{IEEEproof}

From Theorem \ref{thu}, after rounding $\left(M_m+1\right)u_m^*$ and $\left(M_s+1\right)u_s^*$, we can obtain the optimal $\tilde{K}_m^*,\tilde{K}_s^*$ that maximizes $\tilde{T}$. Interestingly, although it is obvious that $\tilde{K}_m^*,\tilde{K}_s^*$ depends on $\frac{\lambda_s}{\lambda_m}$, $\frac{P_s}{P_m}$ and $B$, we find that $\tilde{K}_m^*,\tilde{K}_s^*$ are insensitive to these parameters in numerical results.  The optimality of $\tilde{K}_m^*,\tilde{K}_s^*$ for the original expression $T$  will be also validated through numerical results.

\subsection{Numerical Illustrations}
\begin{figure}[!t]
\centering
\includegraphics[width=3.1in]{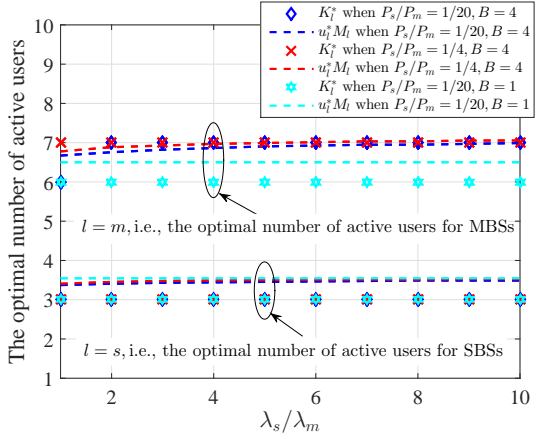}
\caption{The optimal number of active users. $M_m=10$, $M_s=5$, $\alpha=4$.}
\label{OptK}
\end{figure}

By exhaustive  search, we obtain the optimal $K_m^*,K_s^*$ that maximizes $T$. Fig. \ref{OptK} illustrates $K_m^*,K_s^*$ under different system parameters.  It is obvious that $K_m^*$ is either $\lfloor u_m^*\left(M_m+1\right) \rfloor$ or $\lceil u_m^*\left(M_m+1\right) \rceil$, $K_s^*$ is either $\lfloor u_s^*\left(M_s+1\right) \rfloor$ or $\lceil u_s^*\left(M_s+1\right) \rceil$. That is to say, the solution based $\tilde{T}$ can be applied to $T$ directly. Interestingly, we find $u_m^*,u_s^*$ is always located in $[0.59, 0.64]$ even under different system parameters. Therefore, the optimal $K_m^*,K_s^*$ are insensitive to different configurations between the two tiers. Specifically, $K_m^*=6 \textrm{ or } 7$ and $K_s^*=3$ for different $\frac{\lambda_s}{\lambda_m}$, $\frac{P_s}{P_m}$, $B$.

\begin{figure}[!t]
\centering
\includegraphics[width=3.1in]{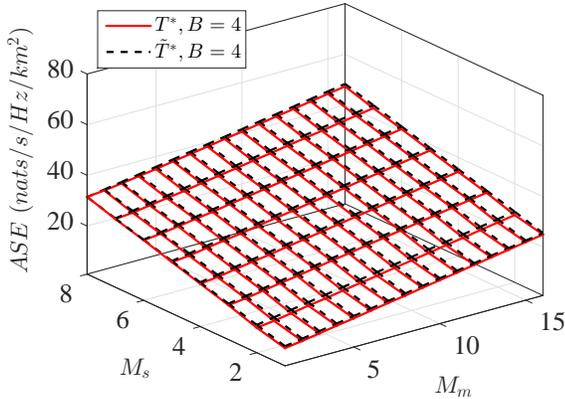}
\caption{$T^*,\tilde{T}^*$ with respect to $M_m,M_s$. $\frac{\lambda_s}{\lambda_m}=5$, $\frac{P_s}{P_m} = \frac{1}{20}$, $\alpha=4$.}
\label{T}
\end{figure}

When the number of active users set as optimal, Fig. \ref{T} depicts $T,\tilde{T}$ with respect to $M_m,M_s$. We find the mismatch between $\tilde{T}^*$ and $T^*$ is quite small, which demonstrates the tightness of Theorem \ref{thlowerbound}. As we mentioned in subsection A, with optimal $u_m^*$ and $u_s^*$, $\tilde{T}$ will increase linearly with $M_m,M_s$. Consistent with $\tilde{T}^*$, we can see that $T^*$ also increases linearly with  $M_m,M_s$. The reasons for this phenomenon are as follows. From Theorem \ref{thu}, we know that $K_m^*,K_s^*$ are approximately fixed portion of $M_m+1,M_s+1$, respectively. In such situation, from Theorem \ref{thlowerbound}, we know that $R_m,R_s$ nearly remain fixed for different $M_m,M_s$. However, the number of active users $K_m^*,K_s^*$ will increase linearly with $M_m,M_s$. In summary, $T^*$ will increase linearly with $M_m,M_s$.

\begin{figure}[!t]
\centering
\includegraphics[width=3.1in]{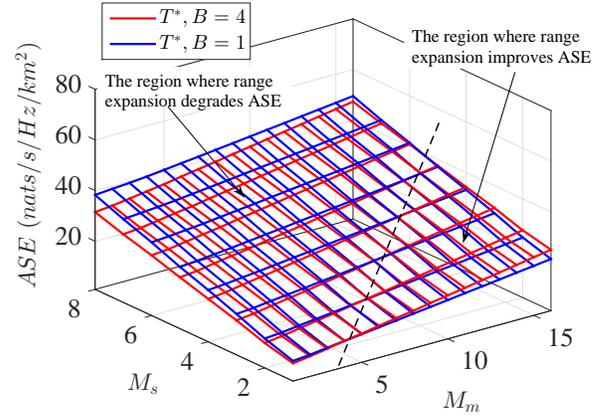}
\caption{$T^*$ with respect to $M_m,M_s,B$. $\frac{\lambda_s}{\lambda_m}=5$, $\frac{P_s}{P_m} = \frac{1}{20}$, $\alpha=4$.}
\label{T14}
\end{figure}

\begin{figure}[!t]
\centering
\includegraphics[width=3.1in]{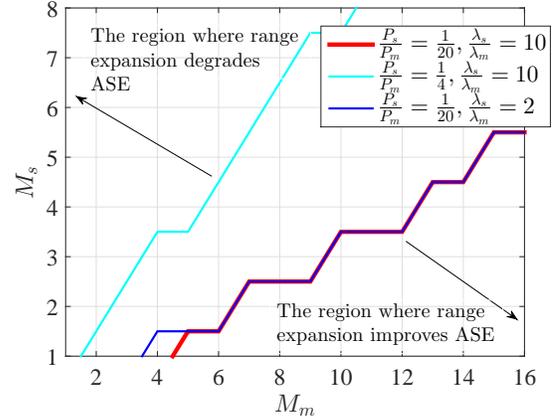}
\caption{The range expansion improvement/degradation region. $\alpha=4$.}
\label{region}
\end{figure}

With the optimal $K_m^*,K_s^*$, Fig. \ref{T14} depicts $T^*$ with $M_m,M_s$ considering different range expansion bias $B$. In single-antenna HetNets, it is found that range expansion will degrade the overall ASE \cite{hetnet:jo}. However, it is not true for multi-antenna HetNets. From Fig. \ref{T14}, if $M_m$ is sufficiently larger than $M_s$, range expansion will improve the ASE. The reasons are as follows. As we mentioned in Theorem \ref{thlowerbound}, range expansion will improve $R_m$ but degrade $R_s$. If $M_m$ is sufficiently larger than $M_s$, the value of $R_m$ will be sufficiently larger than $R_s$. Hence, $R_m$ will benefit more from range expansion. Besides, as $K_m^* \approx u_m^*\left( M_m+1\right)$, the larger $M_m$ is, the more macro users will benefit from the range expansion. Thus, the improvement of $\lambda_mK_mR_m$ due to range expansion will make up for the degradation of $\lambda_sK_sR_s$. Fig. \ref{region} illustrate the regions where range expansion improves or degrades ASE under different system parameters. It is interesting to point out that, the improvement/degradation region is insensitive to $\frac{\lambda_s}{\lambda_m}$. Instead, the difference in transmit power $\frac{P_s}{P_m}$ significantly affects the improvement/degradation region. Specifically, the large $\frac{P_s}{P_m}$ is, the improvement region will be larger. Based on Theorem \ref{thlowerbound}, we know that increasing  $\frac{P_s}{P_m}$ improves both $R_m,R_s$. However, from   Fig. \ref{region}, we know that the improvement of $\lambda_mK_mR_m$ is larger than that of $\lambda_sK_sR_s$, which leads to the fact that the improvement region will be expanded with the increase of $\frac{P_s}{P_m}$.

\section{Conclusions}\label{secconclusion}
In this paper, we have studied the users' average data rate and the optimal settings in multi-antenna two-tier networks. With the help of the tractable approximations, the key properties of system performance have been revealed. We find that the users' data rate of each tier is insensitive to the number of active users of the other tier. If $B>1$, we find users' data rate increases with $\frac{\lambda_s}{\lambda_m},\frac{P_s}{P_m}$. However, if $B=1$, the users' data rate is insensitive to  $\frac{\lambda_s}{\lambda_m},\frac{P_s}{P_m}$. For the purpose of ASE maximization, we find the optimal number of active users of each tier is fixed portion of the sum of the number of antennas plus one. The optimal settings are insensitive to different system parameters. With optimal settings, ASE will increase linearly with the number of antennas. Moreover, we find that, if the number of antennas of MBSs is sufficiently larger than SBSs, range expansion will improve ASE. The improvement region will expand with larger $\frac{P_s}{P_m}$.

This work has revealed the key properties of multi-antenna two-tier networks. Further extension of this work is to consider more sophisticated multi-antenna techniques, such as coordinated beamforming and transmit-receive diversity.

\appendix

\begin{IEEEproof}[\bf{Proof of Theorem \ref{thdatarate}}]We follow the similar approach with \cite{datarate:renzo}. However, we obtain much more simple expressions for multi-antenna HetNets. Take $R_m$ as an example. $E[\log(1+SIR_m)]=\int_0^{\infty}f_{m}(r)E\left[\log \left(1+SIR_m\right)|r\right]dr$, where $f_{m}(r)={2\pi r\left(\lambda_m+\lambda_s\left(\frac{P_sB}{P_m}\right)^{2/\alpha}\right)}e^{-\lambda_m \pi r^2 - \lambda_s \pi \left(\frac{P_sB}{P_m}\right)^{\frac{2}{\alpha}}r^2}$ is the distance distribution between the macro user and the serving MBS $x_0$ \cite{hetnet:jo}. Following the lemma in \cite{hamdi:lemma},
$\begin{array}{l}
E\left[\log \left(1+SIR_m\right)|r\right]=\int_0^{\infty}\frac{1-\mathcal{L}_{g_{m0}}\left(z\right)}{z}\mathcal{L}_{\bar{I}_m}\left(z\right)\mathcal{L}_{\bar{I}_s}\left(z\right)dz
\end{array}$, where $\bar{I}_m={\sum_{x_i  \in \Phi _m \backslash \{ x_0 \} } \frac{h_{mi}r^{\alpha}}{\left\| {x_i } \right\|^{ \alpha }}  } $ and $\bar{I}_s={\sum_{x_i  \in \Phi _s } {\frac{P_s K_mh_{mi}r^{\alpha}}{P_m K_s\left\| {x_i } \right\|^{  \alpha } } } }$.
As $g_{m0}$ is Gamma distributed, we have $\mathcal{L}_{g_{m0}}\left(z\right)=\left(1/(1+z)\right)^{M_m+1-K_m}$. From the properties  of PPP, we have
$\begin{array}{l}
\mathcal{L}_{\bar{I}_m}\left(z\right)=e^{ - \lambda _m \pi r^2 \left(F(z,K_m)-1\right)   }\end{array}$,
$\begin{array}{l}
\mathcal{L}_{\bar{I}_s}\left(z\right)=e^{ - \lambda _s \pi  r^2 \left( {\frac{{P_sB }}{{P_m }}} \right)^{2/\alpha } \left(F\left(\frac{zK_m}{K_sB},K_s\right)-1\right)  }
\end{array}$. Above all, substituting $E[\log(1+SIR_m)|r]$ into the primary integral, we can derive $R_m$ after some algebraic manipulations.
\end{IEEEproof}

\begin{IEEEproof}[\bf{Proof of Theorem \ref{thlowerbound}}]The main procedures are similar to Theorem \ref{thdatarate}, except that we derive an approximation for $E[\log(1+SIR_m)|r]$ by averaging the equivalent channel gains $g_{m0},h_{mi}$ and $h_{si}$. Specifically, we have
\begin{equation}
\begin{array}{l}
E[\log(1+SIR_m)|r] \\
\approx E_{\Phi_m,\Phi_s}\left[\log\left(1+\frac{\frac{P_mE\left[g_{m0}\right]}{K_m r^{\alpha}}}{\sum \limits_{x_i\in \Phi_m \backslash \{x_0\}}\frac{P_mE\left[h_{mi}\right]}{K_m\|x_i\|^{\alpha}}+\sum \limits_{x_i\in \Phi_s }\frac{P_sE\left[h_{si}\right]}{K_s\|x_i\|^{\alpha}}}\right)\right]\\
= E_{\Phi_m,\Phi_s}\left[\log\left(1+\frac{\frac{M_m+1-K_m}{K_m}}{\sum \limits_{x_i\in \Phi_m \backslash \{x_0\}}\frac{r^{\alpha}}{\|x_i\|^{\alpha}}+\sum \limits_{x_i\in \Phi_s }\frac{P_sr^{\alpha}}{P_m\|x_i\|^{\alpha}}}\right)\right] \\
= \int_0^{\infty}\frac{1-e^{-z\frac{M_m+1-K_m}{K_m}}}{z} \\
\quad \quad  \quad \quad \quad e^{-\lambda_m\pi r^2\left(H(z)-1\right)-\lambda_s\left(\frac{P_sB}{P_m}\right)^{2/\alpha}\pi r^2 \left(H(z/B)-1\right) }dz.
\end{array}
\end{equation}
\end{IEEEproof}

\begin{IEEEproof}[\bf{Proof of Lemma \ref{lemmadatarate}}]First consider the case $B>1$. We first show $\tilde{R}_m$ increases with $\frac{\lambda_s}{\lambda_m}$ and $\frac{P_s}{P_m}$. Let $t=\frac{\lambda _s}{\lambda _m} (\frac{{P_s}}{{P_m }})^{\frac{2}{\alpha }}$. Based on Leibniz integral rule, we only need to prove the integral item $\begin{array}{l}{\frac{{{1  + \frac{\lambda _s}{\lambda _m} (\frac{{P_s B}}{{P_m }})^{2/\alpha} } }}{{ H\left( z \right) +  \frac{\lambda _s}{\lambda _m} \left( {\frac{{P_s B}}{{P_m }}} \right)^{2/\alpha} H\left( {\frac{z}{B}} \right)}}}\end{array}$ increases with $t$. Take the first order derivative, we have $\frac{\partial}{\partial t}\left({\frac{{{1  +  B^{2/\alpha}t} }}{{ H\left( z \right) +  t B^{2/\alpha} H\left( {\frac{z}{B}} \right)}}}\right)=\frac{B^{2/\alpha}\left(H(z)-H(\frac{z}{B})\right)}{\left(H\left( z \right) +  t B^{2/\alpha} H\left( {\frac{z}{B}} \right)\right)^2}$. We find $H(x)$ increases with $x$, i.e., $H(x) > H(\frac{x}{B})$. Thus, $\frac{\partial}{\partial t}\left({\frac{{{1  +  B^{2/\alpha}t} }}{{ H\left( z \right) +  t B^{2/\alpha} H\left( {\frac{z}{B}} \right)}}}\right)>0$. Therefore,  $\tilde{R}_m$ increases with $\frac{\lambda_s}{\lambda_m}$ and $\frac{P_s}{P_m}$. Similarly, we can prove  $\tilde{R}_s$ increases with $\frac{\lambda_s}{\lambda_m}$ and $\frac{P_s}{P_m}$. Then, we will prove $R_m$ increase with $B$. Indeed, we have $\begin{array}{l}\frac{\partial}{\partial B}\left({\frac{{{1  +  B^{\frac{2}{\alpha}}t} }}{{ H\left( z \right) +  t B^{\frac{2}{\alpha}} H\left( {\frac{z}{B}} \right)}}}\right)=\frac{2tB^{\frac{2}{\alpha}-1}}{\alpha}\frac{H(z)-e^ {-\frac{z}{B}}+tB^{\frac{2}{\alpha}}\left(H\left(\frac{z}{B}\right)-e^ {-\frac{z}{B}}\right)}{\left({{ H\left( z \right) +  t B^{\frac{2}{\alpha}} H\left( {\frac{z}{B}} \right)}}\right)^2}\end{array}$. It is obvious that $H(x)>1$ and $e^{-\frac{z}{B}}<1$. Thus, $\begin{array}{l}\frac{\partial}{\partial B}\left({\frac{{{1  +  B^{\frac{2}{\alpha}}t} }}{{ H\left( z \right) +  t B^{\frac{2}{\alpha}} H\left( {\frac{z}{B}} \right)}}}\right)>0\end{array}$, i.e., $\tilde{R}_m$ increases with $B$. Following similar procedures, we can prove $\tilde{R}_s$ decreases with $B$. For the case $B=1$, the results can be obtained easily by substituting $B=1$.
\end{IEEEproof}

\ifCLASSOPTIONcaptionsoff
  \newpage
\fi

\end{document}